\documentclass{iopart}

\usepackage{graphicx}
\usepackage{color}
\usepackage{cite}
\usepackage{url}

\newcommand{\HFM}{(H$_2$)$_4$CH$_4$}

\begin{document}

\title[H4M in MOFs and nanotubes]{A theoretical study of the
hydrogen-storage potential of \HFM\ in metal organic framework materials
and carbon nanotubes}

\author{Q. Li and T. Thonhauser}

\address{Department of Physics, Wake Forest University, North Carolina,
27109, USA.}

\ead{thonhauser@wfu.edu}

\begin{abstract}
The hydrogen-methane compound (H$_2$)$_4$CH$_4$---or for short
\emph{H4M}---is one of the most promising hydrogen-storage materials.
This van der Waals compound is extremely rich in molecular hydrogen:
33.3~mass\%, not including the hydrogen bound in CH$_4$; including it,
we reach even 50.2~mass\%.  Unfortunately, H4M is not stable under
ambient pressure and temperature, requiring either low temperature or
high pressure. In this paper, we investigate the properties and
structure of the molecular and crystalline forms of H4M, using \emph{ab
initio} methods based on van der Waals DFT (vdW-DF). We further
investigate the possibility of creating the pressures required to
stabilize H4M through external agents such as metal organic framework
(MOF) materials and carbon nanotubes, with very encouraging results. In
particular, we find that certain MOFs can create considerable pressure
for H4M in their cavities, but not enough to stabilize it at room
temperature, and moderate cooling is still necessary. On the other hand,
we find that all investigated carbon nanotubes can create the high
pressures required for H4M to be stable at room temperature, with direct
implications for new and exciting hydrogen-storage applications.
\end{abstract}

\pacs{71.15.Mb, 84.60.Ve, 88.30.R-, 61.48.De}

\submitto{\JPCM}

\section{Introduction}

The use of hydrogen as an environmentally clean and efficient fuel for
mobile applications is a very active research area. But, before a future
hydrogen economy can become reality, several crucial challenges need to
be addressed, mostly concerning the storage of hydrogen itself
\cite{DOE_report_2}.  The most important challenges for practical
hydrogen-storage system are: (i) high gravimetric and volumetric storage
density (fuel tanks should be light and small); (ii) good thermodynamics
(the hydrogen adsorption/desorption should occur at a reasonable
pressure and temperature); and (iii) fast reaction kinetics (tank
emptying and refilling should be fast). According to the Department of
Energy, addressing these key elements through fundamental research is
imperative to achieving a practical hydrogen economy
\cite{DOE_report_1}. In this paper we present results that address the
gravimetric and volumetric storage density.

A wide variety of materials have been considered as possible
hydrogen-storage materials \cite{Zuttel_04, Crabtree_04, Lim_10,
Sakintuna_07} (see Ref.~\cite{Makowski_09} for a graphical
representation of how many articles have been written on the subject).
However, amongst all these materials, \HFM---or for short H4M---shows
exceptional promise \cite{Mao07}; the H4M system is extremely rich in
molecular hydrogen, containing 33.3 mass\% molecular hydrogen, not
counting the atomic hydrogen in CH$_4$; including it, we reach even
50.2~mass\%.  Simply said, H4M holds more hydrogen per mass and volume
than any known material except pure hydrogen itself! Unfortunately, H4M
is not stable under ambient pressure and temperature.  The stability
field is reported from approximately 5.8~GPa at room temperature to 10~K
for ambient pressure \cite{Mao05, Somayazulu96}, and it has been shown
that moderate cooling can reduce the required pressure significantly
\cite{Mao04}.  This fact led us to design a novel host+H4M structure in
which the host material provides the necessary pressure for H4M to be
stable, without the need for excessive cooling. The use of a host
material will lower the exceptional volumetric and gravimetric hydrogen
storage density of H4M---it is the main goal of this paper to quantify
the tradeoff between lower storage density and stability at
closer-to-ambient temperature and explore if it is worth pursuing.

Since H4M is stable at room temperature under a pressure of 5.8~GPa, we
are most interested in this region of the phase diagram and throughout
the manuscript we report results at that pressure. We first consider H4M
in nanoporous metal organic framework (MOF) materials \cite{Yaghi95,
Rosi03, Zhao04, Eddaoudi02}, which provide significant pressure inside
their cavities to stabilize H4M. We study MOFs that are isoreticular to
(i.e.\ have the same network topology as) MOF-5, with varying linker
length. We find that these MOFs all provide enough pressure to
significantly decrease the burden of cooling, but none of them stabilize
H4M at room temperature. As a second host material, we consider
single-wall carbon nanotubes (CNT)\cite{Eklund}, which are well known
for being able to provide high pressure inside their cavities.
Specifically, we study zigzag nanotubes with chirality from (10,0) to
(26,0) and corresponding radii from 4~\AA~to 10~\AA---while larger
nanotubes with a radius up to several nanometers can be produced in
experiments \cite{Cheung02}, such systems together with H4M filling
inside are, at the moment, not accessible through \emph{ab initio}
simulations. Our calculations show that crystalline H4M may be
stabilized at room temperature inside nanotubes, opening the door for
high-efficiency hydrogen-storage applications.

This paper is organized as follows: After presenting computational
details in section \ref{sec:comp}, we show results for the structure of
H4M in its molecular and crystalline form in section
\ref{sec:structure}. Results for H4M inside the cavities of MOFs and
nanotubes are presented in section \ref{sec:H4M_in_MOF} and
\ref{sec:H4M_in_CNT}, respectively. We conclude and suggest future
research in section \ref{sec:conclusions}.

\section{Computational Details}
\label{sec:comp}

H4M is classified as a van der Waals compound \cite{Mao05}. The study of
nanotube interactions \cite{Soler_09} and binding of small molecules in
MOFs \cite{MOF_timo} is also strongly determined by van der Waals
interactions. As such, an accurate description of van der Waals
interactions in these systems is crucial.  Thus, we use the recently
developed van der Waals density functional (vdW-DF), a truly nonlocal
exchange and correlation functional, that incorporates van der Waals
forces self-consistently and seamlessly into DFT \cite{Dion05,
Thonhauser07, Langreth09}. The vdW-DF approach has shown good
transferability for a range of van der Waals systems reaching from
simple dimers \cite{dimers} and physisorbed molecules
\cite{physisorption} to DNA \cite{DNA} and drug design \cite{drug}. In
particular, it has been applied successfully to answer questions
regarding hydrogen storage \cite{clathrates}, MOFs \cite{MOF_timo}, and
nanotubes \cite{Soler_09}.

For our calculations we use density functional theory, as implemented in
the \textsc{PWscf} code, which is part of the \textsc{Quantum-Espresso}
package \cite{Giannozzi09}.  We utilize ultrasoft pseudopotentials with
a wave-function cutoff of 475~eV (5700~eV charge-density cutoff) for
carbon nanotube related systems and 475~eV (3800~eV charge-density
cutoff) for all other cases.  A self-consistency convergence criterion
of at least 1.4 $\times$ 10$^{-7}$~eV is used. All structures are
relaxed until all force components are less than 2.5 meV/\AA.  Due to
the limitation of calculation size, and the fact that larger MOFs
provide less pressure anyway, only the smallest MOF structures are
investigated, using a $2\times 2\times 2$ Monkhorst-Pack {\bf k}-mesh
\cite{MONKHORST_PACK}. For simulations of crystalline H4M itself, we use
a $4\times 4\times 4$ {\bf k}-mesh.  All carbon nanotube related systems
are simulated as infinitely long hexagonal nanotube arrays with varying
radius/distance, comprised of a short unit cell length in $z$-direction,
using a $1\times 1\times 16$ {\bf k}-mesh. Single molecule calculations
on H4M molecules are performed in a $16\times 16\times 16$ \AA$^3$ cubic
unit cell.

\section{Structure of H4M}
\label{sec:structure}

Although H4M under pressure has been investigated by X-ray diffraction
\cite{Somayazulu96}, the actual microscopic structure of crystalline H4M
remains unknown \cite{Mao05}. The X-ray diffraction experiments suggest
a body-centered tetragonal structure \cite{Somayazulu96} and Raman
experiments provide further constraints \cite{Mao05}. Here, starting
from these experimental guidelines, we report results for \emph{ab
initio} calculations of the molecular and crystalline H4M structure. We
use a bottom-up approach, where we first find the optimal structure for
H4M molecules, which we then assemble to form solids with various
symmetries.

\subsection{Structure of molecular H4M}

\begin{figure}
\begin{indented}\item[]
\includegraphics[width=0.3\columnwidth]{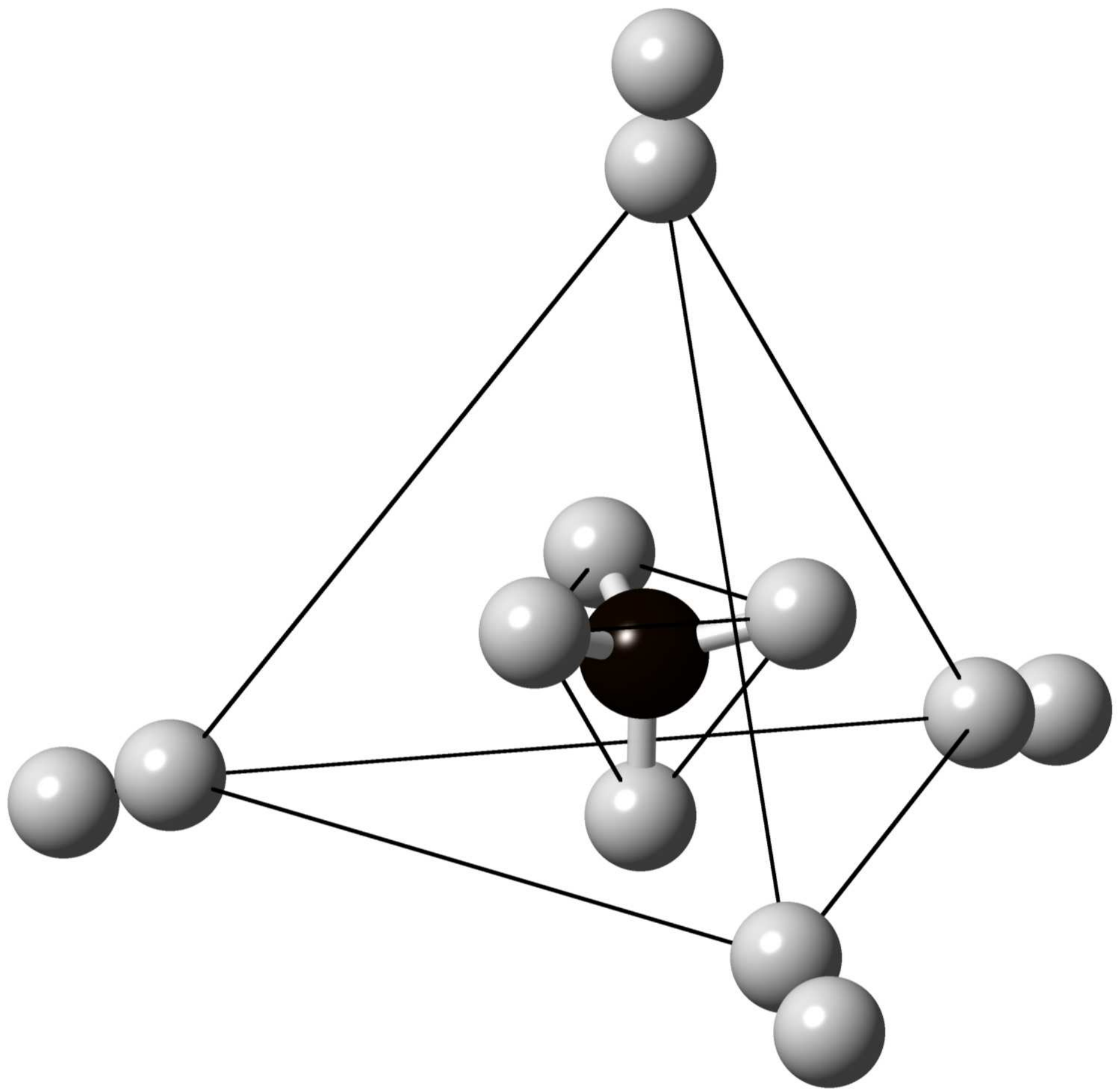}
\caption{\label{fig:h4m_molecule}Optimized structure of a single H4M
molecule \HFM. The methane molecule forms a small tetrahedra, which sits
inside a large tetrahedra formed by the hydrogen molecules. The axes of
all hydrogen molecules point toward the centers of faces of the methane
tetrahedra.}
\end{indented}
\end{figure}

We start by investigating the structure of a single H4M molecule. After
relaxing the molecule, starting from a variety of initial positions, we
always find the same optimal structure: the molecule forms two
tetrahedra, sharing the same center with opposite directions. The common
center is the carbon atom of the methane molecule, which forms the small
tetrahedron. The four hydrogen molecules form a larger tetrahedron, each
of them located on one of the vertexes, and oriented so that they point
to the central carbon atom of the methane. We find the C--H bond of the
methane molecule to be 1.091~\AA, the H--H bond of the hydrogen
molecules to be 0.742~\AA, and the distance from the central carbon atom
to the closer hydrogen atom of the hydrogen molecules to be 3.295~\AA.
The optimized structure of the H4M molecule is depicted in
\fref{fig:h4m_molecule}.

In \tref{tab:H2_binding_energies}, we list the binding energy as a
function of the number of H$_2$ molecules bound to the central CH$_4$
molecule. Here, the binding energy is defined as $E_{\rm bind} = E[({\rm
H}_2)_n{\rm CH}_4] - E[{\rm CH}_4] - n\,E[{\rm H}_2]$ for the case of
$n$ H$_2$ molecules bound to the CH$_4$ molecule. We can see that an
increase in the number of H$_2$ molecules leads to an increase in
binding energy per H$_2$ molecule, due to the additional interaction of
H$_2$ molecules.  We also report the incremental absorption energy,
i.e.\ the work to absorb each new H$_2$ molecule. The increasing
magnitude in both columns validates the stability of the H4M molecule.

\begin{table}
\caption{\label{tab:H2_binding_energies} Binding energies
$E_{\rm{bind}}$ [meV] of H$_2$ in (H$_2$)$_n$CH$_4$ as a function of the
number of H$_2$ molecules attached. We also report the incremental
absorption energy $E_{\rm abs}^{\rm inc}$ [meV], i.e.\ the work to
absorb each new H$_2$ molecule.}
\begin{indented}\item[]
\begin{tabular}{@{}lccr@{}}
\br
$n$ & $E_{\rm{bind}}$ & $E_{\rm{bind}}$/H$_2$ & $E_{\rm abs}^{\rm inc}$\\
\mr
1 & --19.50 & --19.50 & --19.50\\
2 & --39.58 & --19.79 & --20.08\\
3 & --60.33 & --20.11 & --20.75\\
4 & --81.36 & --20.34 & --21.03\\
\br
\end{tabular}
\end{indented}
\end{table}

At this point, it is instructive to pause for a moment and analyze the
performance of vdW-DF in H4M, the structure and binding of which is very
much determined by van der Waals forces. Since this molecule is small
enough, we can compare here with the highest level of quantum chemistry.
To this end, we have calculated the binding energy of four H$_2$
molecules to methane---i.e.\ the last row of
\tref{tab:H2_binding_energies}---using various approaches and
exchange-correlation functionals. As reference, we use MP2 and CCSD(T)
calculations, using the aug-cc-pVDZ basis set with counterpoise
corrections.  For the binding energy we find --85.93~meV with MP2 and
--87.75~meV with CCSD(T).  These reference numbers are now to be
compared with DFT calculations. Using standard LDA \cite{LDA}, we find a
very strong erroneous overbinding of --194.37~meV, not untypical when
applying LDA to small molecules with van der Waals binding \cite{Brian}.
On the other hand, a standard PBE \cite{PBE} calculation gives a large
erroneous underbinding of only --36.65~meV. From these results it is
clear that a simple H4M molecule is a very delicate system that is very
much influenced by van der Waals interactions and thus an ideal
application for vdW-DF. The corresponding binding energy with vdW-DF is
--81.36~meV, in almost perfect agreement with high-level quantum
chemistry. These results show that standard functionals such as LDA and
PBE are not suited to describe the binding of H4M itself and much less
so to study the H4M crystal structure or its properties inside MOFs and
nanotubes; thus, in the following we will only report results for
vdW-DF.

Also worth mentioning at this point is that vdW-DF is responsible for a
small but significant charge transfer; in fact, it seems that vdW-DF is
self-consistently rearranging the charge density just right, resulting
in the excellent binding energy. Such charge transfers caused by vdW-DF
are typically very small, i.e.\ on the order of 10$^{-4}$
electrons/\AA$^3$, where electronic charge gets accumulated between the
constituents. A nice presentation of this charge transfer due to van der
Waals binding is given in figure~8 of Ref.~\cite{Thonhauser07}.

\subsection{Structure of crystalline H4M}

To find the true crystalline structure for H4M is a complicated task.
In principle, one should use an approach such as the \emph{ab initio}
random structure search (AIRSS) algorithm \cite{Pickard_11}---which
generates random structures biased on chemistry, experimental, and
symmetry grounds---to sample the corresponding large phase space. Here,
however, we make use of experimentally suggested symmetries
\cite{Somayazulu96, Mao05} and an already existing semiclassical
sampling of a large number of possible configurations \cite{Maoppt}.

After Somayazulu et al.\ suggested a body-centered tetragonal (BCT)
structure (with some uncertainty) based on X-ray diffraction patterns
\cite{Somayazulu96}, research on the precise structure of crystalline
H4M seems to have subsided. Fifteen years later, Mao et al.\ found a
simple orthorhombic unit cell for crystalline H4M from a combined
semiclassical/DFT approach by randomly arranging atoms, which resulted
in a good agreement between calculated and experimental XRD spectra
\cite{Maoppt}. They found an optimized unit cell of 332~\AA$^3$ with
four H4M molecules, resulting in a molecular hydrogen volumetric density
of 0.16~kg H$_2$/L, in disagreement with their previously published
number, which is almost twice as large \cite{Mao07}---which we thus
believe to be an error.

\begin{figure}
\begin{indented}
\item[]\includegraphics[width=0.6\columnwidth]{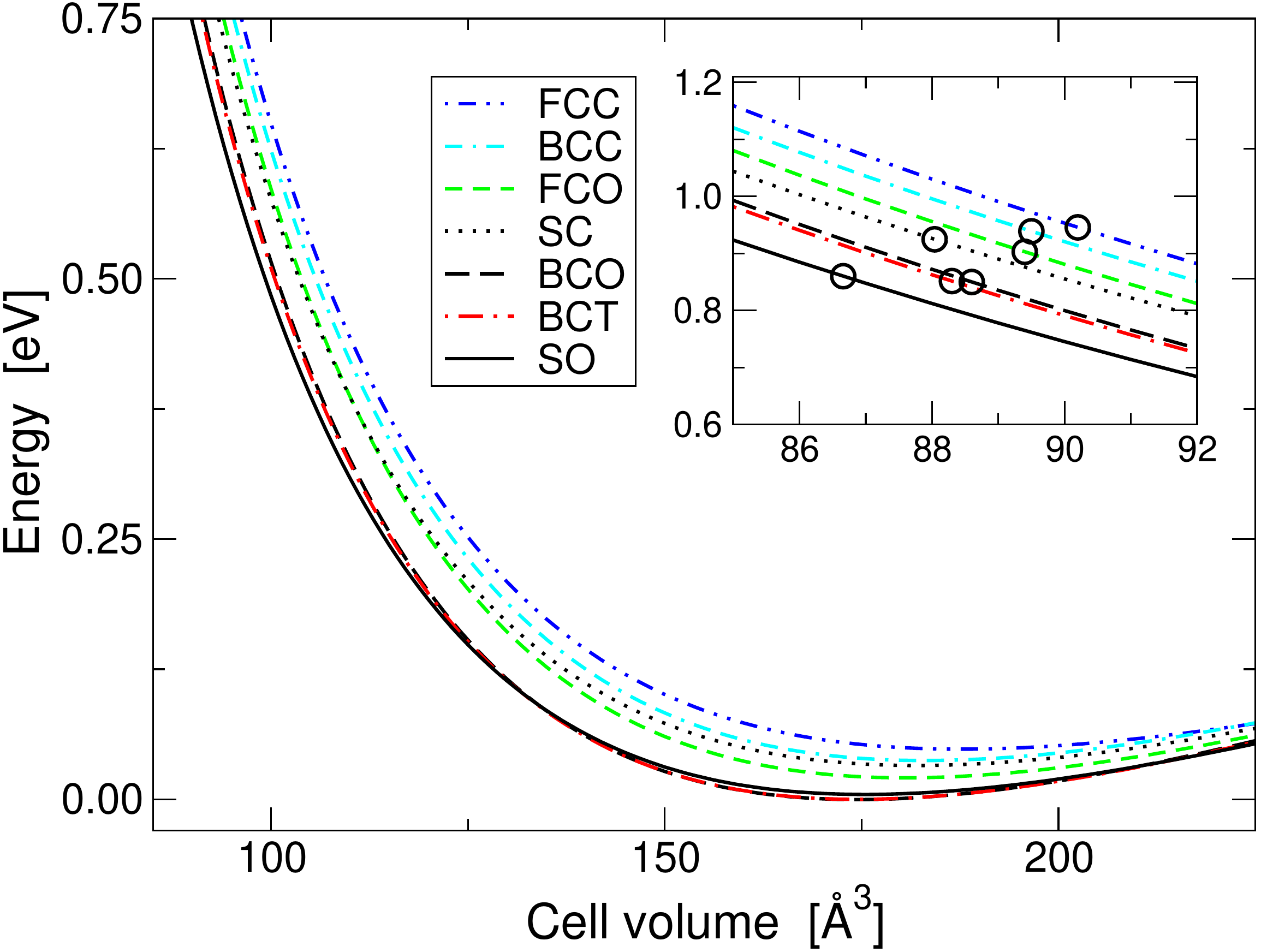}
\caption{\label{fig:H4Mcrystal}Energy versus volume curve for seven
possible symmetries of the H4M crystal. The circles in the small inset
indicate the 5.8 GPa point of each structure. Energies are plotted
relative to the SO energy at zero pressure.}
\end{indented}
\end{figure}

Using the above information as starting point, we generate seven closely
related possible unit cells with an H4M molecule as a building block
(using the molecular structure described above), i.e.\
simple/body-centered/face-centered cubic (SC/BCC/FCC),
simple/body-centered/face-centered orthorhombic (SO/BCO/FCO), and
body-centered tetragonal (BCT). For all seven structures, we internally
(atom positions) and externally (unit cell parameters) optimize the unit
cells and determine the optimal ratios between the lengths of their
edges ($b/a$ and $c/a$) without pressure; the corresponding results are
listed in \tref{tab:lattice_constants}. Then, we add hydrostatic
pressure by changing the cell parameter $a$, while keeping $b/a$ and
$c/a$ constant, and relaxing the atom positions again for each volume.
Applying the exact conditions of hydrostatic pressure requires knowledge
of all corresponding elastic constants, but test calculations show that
H4M is close to isotropic, such that the uniform scaling of all lattice
constants is a good approximation\footnote{E.g.\ the SO unit cell is
almost tetragonal---see \tref{tab:lattice_constants}---and our
calculations show that the corresponding elastic constants c$_{22}$ and
c$_{33}$ differ only by a few percent, while c$_{33}$ differs by about
10\%. The numbers are similar for other symmetries.}. The resulting
energy versus volume curves are fitted using a Murnaghan equation of
state, enabling us to analyze the structures and their pressure
dependence in detail; results are shown in \fref{fig:H4Mcrystal}. As can
be seen, the BCO and BCT curves are very close to each other due to the
similarity of their structures, and they have the lowest energies at
zero pressure.  However, as the pressure increases, they cross with the
SO curve, indicating a pseudo-phase transformation at around 1.8~GPa.
At 5.8 GPa---the pressure at which H4M is experimentally stable at room
temperature---SO has the smallest volume and slightly higher energy than
BCO/BCT. Since at zero temperature the condition for the stable
structure at constant pressure is that enthalpy ($H = E + PV$) be
minimum \cite{MartinBOOK}, we conclude that the SO structure is favored
at 5.8~GPa, in agreement with Mao et al. \cite{Maoppt}. The
corresponding results for zero pressure and a pressure of 5.8~GPa are
listed in \tref{tab:energy_enthalpy}. It is striking to see how ``soft''
this material is, as evident by the tremendous volume change upon
applying 5.8~GPa of pressure, as is expected for a van der Waals
crystal.

\begin{table}
\caption{\label{tab:lattice_constants}Lattice constant $a$ [\AA] and
corresponding ratios $b/a$ and $c/a$ for crystalline H4M in various
symmetries.}
\begin{indented}
\item[]\begin{tabular}{@{}lcccr@{}}
\br
Symmetry & $a$ @  5.8 GPa & $a$ @ 0 GPa & $b/a$ & $c/a$\\
\mr
FCC & 7.119 & 9.089 & 1     & 1    \\
BCC & 5.636 & 7.166 & 1     & 1    \\
FCO & 7.791 & 9.865 & 0.897 & 0.843\\
SC  & 4.449 & 5.684 & 1     & 1    \\
BCO & 6.162 & 7.741 & 0.976 & 0.776\\
BCT & 6.056 & 7.628 & 1     & 0.795\\
SO  & 5.174 & 6.567 & 0.798 & 0.784\\
\br
\end{tabular}
\end{indented}
\end{table}

\begin{table}
\caption{\label{tab:energy_enthalpy} Volume $V$ [\AA$^3$], volumetric
hydrogen density $\rho_{\rm vol}$ [kg H$_2$/L], energy $E$ [meV], and
enthalpy $H$ [meV] of crystalline H4M for various symmetries.  Energies
are reported relative to the values for the SO structure.}
\begin{indented}
\item[]\begin{tabular}{@{}lccccccr@{}}
\br
         & \centre{4}{@ 5.8 GPa}    & \centre{3}{@ 0 GPa}\\
         & \crule{4}                & \crule{3}\\
Symmetry & $V$ & $\rho_{\rm vol}$ & $E$ & $H$ & $V$ & $\rho_{\rm vol}$ & $E$\\
\mr
FCC & 90.19 & 0.1473 &  88.2 & 215.9 & 187.68 & 0.0708 & 42.8 \\
BCC & 89.49 & 0.1484 &  80.5 & 182.9 & 184.01 & 0.0722 & 32.0 \\
FCO & 89.40 & 0.1486 &  44.4 & 143.6 & 181.47 & 0.0732 & 15.4 \\
SC  & 88.04 & 0.1509 &  65.1 & 114.9 & 183.62 & 0.0723 & 27.4 \\
BCO & 88.59 & 0.1499 & --8.1 &  61.7 & 175.63 & 0.0756 & --5.3\\
BCT & 88.31 & 0.1504 & --7.3 &  52.1 & 176.45 & 0.0753 & --5.3\\
SO  & 86.67 & 0.1533 &   0.0 &   0.0 & 177.16 & 0.0750 &  0   \\
\br
\end{tabular}
\end{indented}
\end{table}

\section{H4M in MOFs}
\label{sec:H4M_in_MOF}

\begin{figure}
\begin{indented}
\item[]\includegraphics[width=0.35\columnwidth]{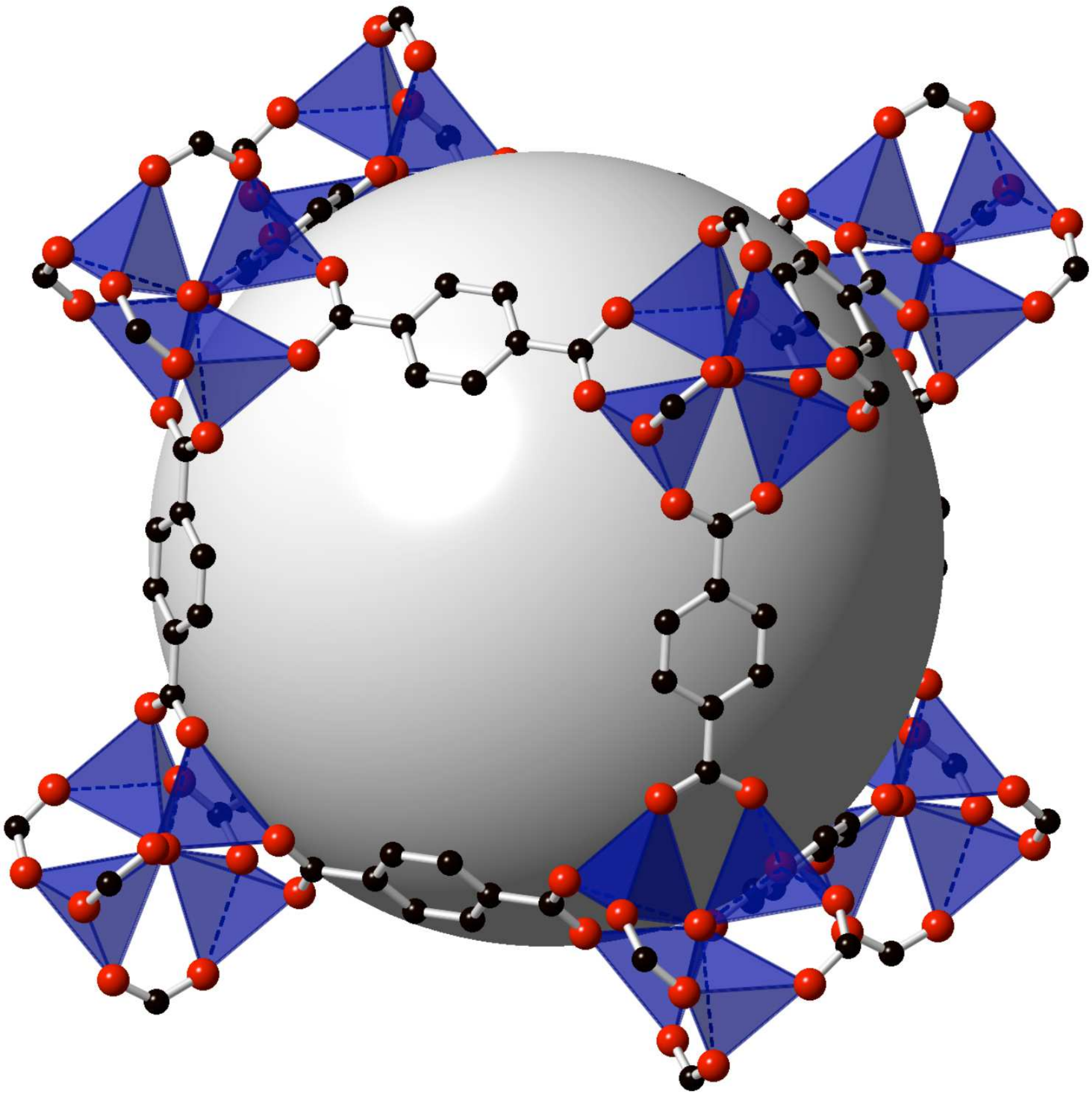}
\caption{\label{fig:MOF-5}Structure of MOF-5. The unit cell consists of
424 atoms (for clarity hydrogen atoms are not displayed). The Zn--O--C
clusters at the corners are connected through one benzene linker.  The
sphere in the middle serves as a measure for the size of the cavity---we
refer to the radius of this sphere as the \emph{fixed radius}.  The
\emph{free radius} is defined as the radius of the largest sphere that
fits through the aperture.}
\end{indented}
\end{figure}

With the H4M crystal structure, we are now ready to position H4M into
external host materials, in the hope to stabilized it at reasonable
temperatures. We begin by investigating the possible pressure range that
several common MOF materials exhibit. For our study, we limit ourselves
to three MOFs of different sizes. The smallest MOF we include is MOF-5,
consisting of Zn--O--C clusters at the corners, connected through one
benzene linker in a cubic symmetry, depicted in \fref{fig:MOF-5}. From
this starting structure, larger MOFs can be build by simply making the
chain of benzene linkers longer, while keeping the network topology the
same. Such MOFs are referred to as \emph{isoreticular}, and we have
selected irMOF-10 (the linker chain contains two benzene rings) and
irMOF-16 (the linker chain contains three benzene rings). For a nice
graphical representation and further details of the three MOFs under
investigation, please see Ref.~\cite{Eddaoudi02}. As the starting point
for the three MOF structures, we use experimental atom positions from
the supplementary materials of Ref.~\cite{Eddaoudi02}.

MOFs can have very large unit cells with hundreds of atoms,
which---together with hundreds of atoms from the H4M filling---can
quickly render \emph{ab initio} calculations impractical. In order to
still be able to make quantitative statements, we begin by investigating
empty MOFs and the pressure they can create in their cavities. With the
knowledge gained about MOF strength, we estimate their performance when
filled with H4M. We verify the results with one full \emph{ab initio}
calculation on a small, but complete system of MOF+H4M.

\begin{figure}
\begin{indented}
\item[]\includegraphics[width=0.6\columnwidth]{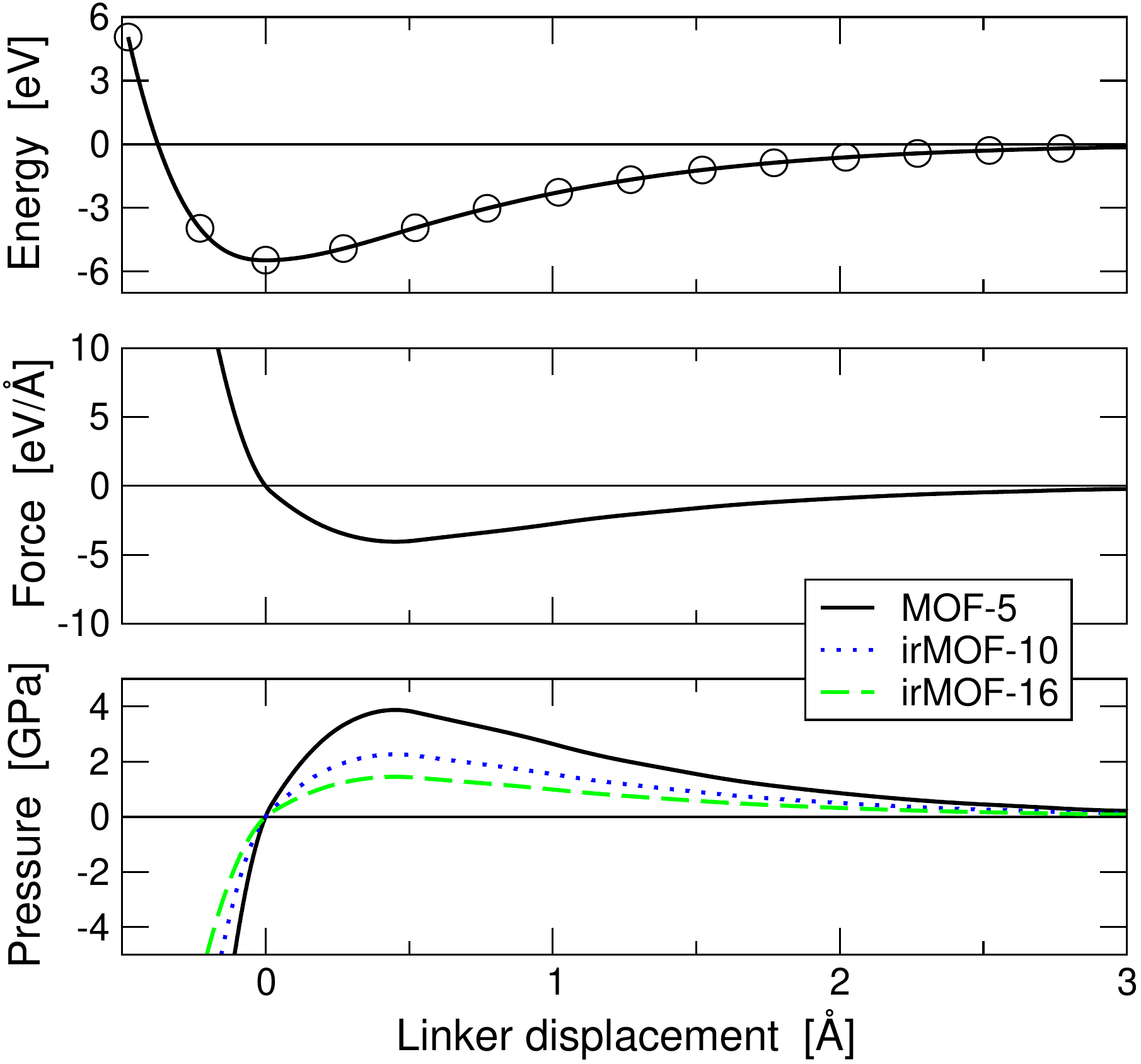}
\caption{\label{fig:MOF_cluster_benzene_bond}Energy and calculated
force/pressure versus benzene ring displacement in $z$-direction of
simulated MOF-5. The energy is plotted relative to the energy of the
completely separated Zn--O--C cluster and benzene ring. In the upper
panel, open circles indicate the \emph{ab initio} results; the solid
lines are spline fittings.}
\end{indented}
\end{figure}

First, we investigate the strength of the bond between the Zn--O--C
cluster corners and the connected benzene linker. Knowing the strength
of this bond allows us to estimate the pressure the MOF can withstand in
its cavity. To this end, we perform total energy calculations in which
we pull the benzene ring away from the cluster. We keep the structure in
$x$ and $y$ directions constant, and move the benzene ring along the
$z$-axis. To accommodate this, we construct a new tetragonal unit cell
with the same cell parameter in $x$ and $y$ directions (12.916 \AA) as
the relaxed cubic MOF-5 structure, but with an elongated $z$-edge with a
$c/a$ value of 2.2. This allows us to move the benzene ring several \AA,
while always keeping a minimum distance of at least 8 \AA\ to the
Zn--O--C cluster in the next unit cell. The side of the benzene that
would usually connect to the next corner was hydrogen terminated.
Throughout the calculations, the relative positions of the atoms in the
Zn--O--C cluster and the benzene rings are fixed, the only element that
changes is the relative distance between the Zn--O--C cluster and the
benzene ring in the $z$-direction. Results of the total energy versus
benzene ring displacement are plotted in the upper panel of
\fref{fig:MOF_cluster_benzene_bond}.  The force is then calculated by
the energy derivative and is plotted in the middle panel of
\fref{fig:MOF_cluster_benzene_bond}. From this, we can estimate the
possible pressure that MOF-5 can create, as the pressure is the ratio
between the force and the area of the square in the cubic unit cell.
Note that the connection of the benzene ring to the Zn--O--C clusters at
the corners is the same for MOF-5, irMOF-10, and irMOF-16. As such, the
energy vs.\ benzene ring displacement plot allows us to estimate the
pressure for all three MOFs.  Simple test calculations show that the
bond between two benzene rings in longer linkers is slighty stronger
than the bond between the Zn--O--C cluster and the first benzene ring.
In other words, the bond between the Zn--O--C cluster and the first
benzene ring breaks before the bond between two benzene linkers. Thus,
the corresponding pressure in irMOF-10 and irMOF-16 can easily be
estimated by dividing by their corresponding (larger) areas. The
pressures, as a function of benzene ring displacement, estimated in this
way are depicted in the lower panel of
\fref{fig:MOF_cluster_benzene_bond}. The zero point on the horizontal
displacement axis indicates the position at which the benzene linker
sits in the optimized zero-pressure unit cell. The maximum pressure in
the case of MOF-5 is 3.9~GPa, close to the required H4M storage pressure
at room temperature.  The larger host materials irMOF-10 and irMOF-16
can still produce a maximum pressure of 2.3~GPa and 1.5~GPa,
respectively, before they break.

We next investigate the possible storage of H4M in these MOF systems.
The free and fixed diameter (see \fref{fig:MOF-5} for a definition of
these terms) of the pores in MOF-5 have been reported as 11.2~\AA\ and
18.6~\AA, respectively \cite{Eddaoudi02}. Since the diagonal length of
our calculated SO H4M structure at 5.8~GPa is 7.7~\AA, we insert a
$2\times 2\times 2$ compressed H4M supercell into the pores of MOF-5.
The MOF-5 structure itself has 424 atoms; fully filled with the H4M
supercell, this corresponds to more than 1000 atoms.  Since this is
computationally too demanding, we only fill one pore (out of eight) per
unit cell, resulting in a 528 atom unit cell. We then perform \emph{ab
initio} calculations on this system and find that the unit cell
experiences a pressure of 3.9~GPa (the fact that this number is the same
as the corresponding maximum pressure in \tref{tab:MOF_performance} is
accidental.) This pressure is lower than the 5.8~GPa compacted H4M
supercell that inserted, simply because the cavity is slightly larger
than the H4M supercell. When we further let all atoms relax again in
this structure, we find that the unit cell stabilizes at approximately
1~GPa. Note that, upon relaxation, we find only minimal changes in the
atoms positions of both the MOF and the inserted H4M.

Based on the estimations from above, in \tref{tab:MOF_performance}, we
list the theoretical performance of H4M storage in MOF-5, irMOF-10, and
irMOF-16. Since the pore size should be approximately commensurate with
the unit cell size of crystalline H4M, we can estimate a possibly
largest H4M supercell that would fit in each pore, according to the pore
diameters \cite{Eddaoudi02}. From that, we calculate the hydrogen mass
density---including and not including the hydrogen in CH$_4$. In
addition, with the knowledge of the experimentally determined stability
field \cite{Mao05} and the possible maximum pressure from above, we
estimate to which temperature the systems need to be cooled for the H4M
to be stable. Overall, there is a trade-off between the cooling required
and mass density---the higher the hydrogen mass density is, the more
cooling is required. Even though none of those three MOFs stabilize H4M
at room temperature, these results are promising: experimentally, pure
hydrogen gas absorption using MOF-5 can reach a maximum mass density of
4~mass\% \cite{Rosi03}, while our calculated hydrogen mass
density---using H4M as a guest molecule---is significantly higher. This
result is related to the fact that most of the pure hydrogen molecules
physisorb on the inside walls of the pores, while using H4M makes more
efficient use of the entire pore volume.

\begin{table}
\caption{\label{tab:MOF_performance} Performance of different MOFs as
hosts for H4M. Given are: MOF type, free and fixed pore diameter [\AA],
maximum commensurate size of the H4M supercell for filling, hydrogen
mass density $\rho_{\rm mass}$ [mass\%] with and without the hydrogen in
CH$_4$, maximum pressure $P$ achievable [GPa], and temperature $T$ [K]
to which systems needs to be cooled to stabilize H4M.}
\begin{indented}
\item[]\begin{tabular}{@{}lccccccr@{}}
\br
System & Free & Fixed & Filling & $\rho_{\rm mass}^{\rm w}$ & $\rho_{\rm mass}^{\rm w/o}$ & Max. $P$ & $T$\\
\mr
MOF-5    & 11.2 & 18.6 & $2\times 2\times 2$ & 10.0 &  6.7 & 3.9 & 224\\
irMOF-10 & 15.4 & 24.5 & $3\times 3\times 3$ & 19.5 & 13.0 & 2.3 & 172\\
irMOF-16 & 19.1 & 28.8 & $3\times 3\times 3$ & 15.3 & 10.2 & 1.5 & 126\\
\mr
Pure H4M & ---  & ---  & ---                 & 50.2 & 33.3 & --- &  10\\ 
\br
\end{tabular}
\end{indented}
\end{table}

\section{H4M in nanotubes}
\label{sec:H4M_in_CNT}

Nanotubes have been studied extensively and it is well known that they
have extraordinary electronic and physical properties \cite{Eklund}. In
particular, nanotubes have exceptional high stiffness and strength and
hence the ability to withstand large elastic strain \cite{Saito,
Yu00PRL}. In the following, we investigate if carbon nanotubes can
provide the necessary pressure to stabilize H4M. We consider zigzag
single-wall carbon nanotubes---since they are semiconducting and thus
easier to model---with chirality ($n,m$) ranging from (10,0) to (26,0),
corresponding to radii from 4~\AA\ to 10~\AA.  While larger nanotubes
with a radius of up to several nanometers can be produced in experiments
\cite{Cheung02}, the filling of such systems with H4M is currently not
accessible through \emph{ab initio} simulations, due to the large number
of atoms. Similar to the case of H4M in MOFs, we first model empty
nanotubes to find their elastic properties and then investigate their
loading with H4M through suitable approximations. Again, we verify our
results by one \emph{ab initio} calculation on a small, but complete
system of CNT+H4M.

\subsection{Radial Young's modulus and strain of carbon nanotubes}

\begin{figure}
\begin{indented}
\item[]\includegraphics[width=0.7\columnwidth]{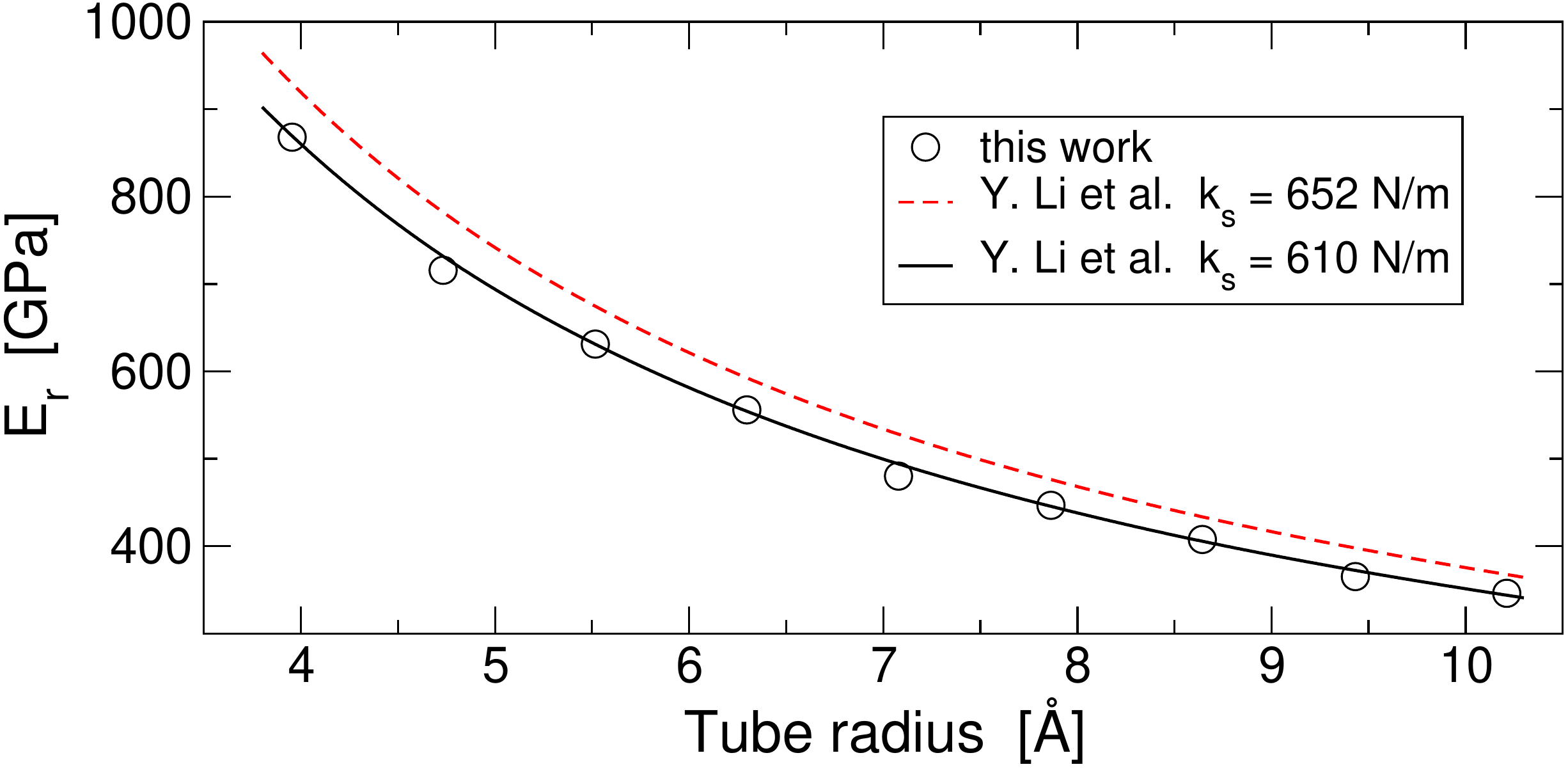}
\end{indented}
\caption{\label{fig:NT_YoungModulus}Calculated radial Young's modulus of
nanotubes of different sizes. {\color{red} }}
\end{figure}

The property of most interest to us is the Young's modulus, as it is
closely related to the pressure resulting from a small perturbation to
the tube's radius. Since the axial Young's modulus has been determined
to be 1 TPa \cite{Lu97, Yu00PRL}, we focus on the radial Young's modulus
here.  We start from the properties of isolated carbon nanotubes,
modeling them with a minimum wall-to-wall separation of at least 8 \AA.
The radial Young's modulus $E_r$ in a hydrostatic pressure model is
given by
\begin{equation}
\Delta U_e = \frac{E_r\,A_0\,\Delta r^2}{2r_0}\;,
\end{equation}
where $\Delta U_e$ is the relative strain energy, $A_0$ and $r_0$ are
the original cross-section and radius, and $\Delta r$ is the amount by
which the radius changes.  For each nanotube, we first relax the
structure and then perform five self-consistent calculations varying the
tube radius by 0\%, $\pm$~0.5\%, and $\pm$~1\%. As expected, $\Delta
U_e$ and $\Delta r$ satisfy a quadratic relation in a small vicinity of
$r_0$, allowing us to calculate $E_r$ for each nanotube. Our results are
depicted in \fref{fig:NT_YoungModulus}, together with recently reported
results from analytical calculations \cite{Li09}, where the following
fit for $E_r$ of zigzag carbon nanotubes has been found:
\begin{equation}
E_r = \frac{2nk_s\,\sin^2(\pi/n)}{3\pi b}\;.
\end{equation}
Here, $n$, $b$, and $k_s$ refer to the chirality $(n,0)$, the C--C bond
length (in our case 1.42~\AA), and the stretching force constant of the
covalent bond. The comparison of our results to the results of Li et al.
\cite{Li09} is already very good, but can be improved if we adjust $k_s$
from its originally published value of 652 N/m to 610 N/m.

Knowledge of the radial Young's modulus allows us to estimate which
pressure the nanotubes can withstand inside its cavity. In particular,
we use the Young's modulus found previously, to calculate the strain
necessary to create 5.8~GPa of pressure inside the nanotube---this is
the pressure required to stabilize H4M at room temperature. The
corresponding strains are given in \tref{tab:H4M-CNT}. Next, we perform
simple calculations on the carbon framework to measure how much strain
it can withstand before it breaks. In our calculations we investigate
strains from 10\% to 20\% and we find that the carbon network breaks
approximately at 15\% strain, in good agreement with Ref.~\cite{Zhao09,
Grantab10}; although other experiments note that breaking occurs at a
lower strain of around 5\% \cite{Yu00PRL, Yu00Science}.

\subsection{Binding between nanotubes, optimized separation for a
nanotube bundle}

We next investigate the binding of nanotubes to each other. To this end
we arrange the nanotubes in a hexagonal unit cell and study the energy
as a function of separation in such bundles. We start from isolated
nanotubes and bring them together incrementally, performing two types of
calculations: (i) the atoms in the tubes are fixed and (ii) the tubes
are free to deform. From the resulting energy curves, we find that
nanotubes bind approximately at a wall-to-wall separation of 3.5~\AA,
nearly independent of size. Furthermore, larger tubes undergo more
deformation at close distance and there is a significant energy gain
from the deformation if the separation is less than 3~\AA\ for tubes
with chirality larger than (16,0). Since the deformation only plays
a significant role in a region closer than the natural binding
distance, we will use
a 3.5~\AA\ wall-to-wall distance and start all simulations from
undistorted nanotubes in the calculations reported below. From our
calculated wall-to-wall separation of nanotubes in bundles, we can then
deduce the achievable volumetric hydrogen-storage density; results are
given in \tref{tab:H4M-CNT}.

\subsection{CNT+H4M system}

\begin{figure}
\begin{indented}
\item[]
\includegraphics[width=0.30\columnwidth]{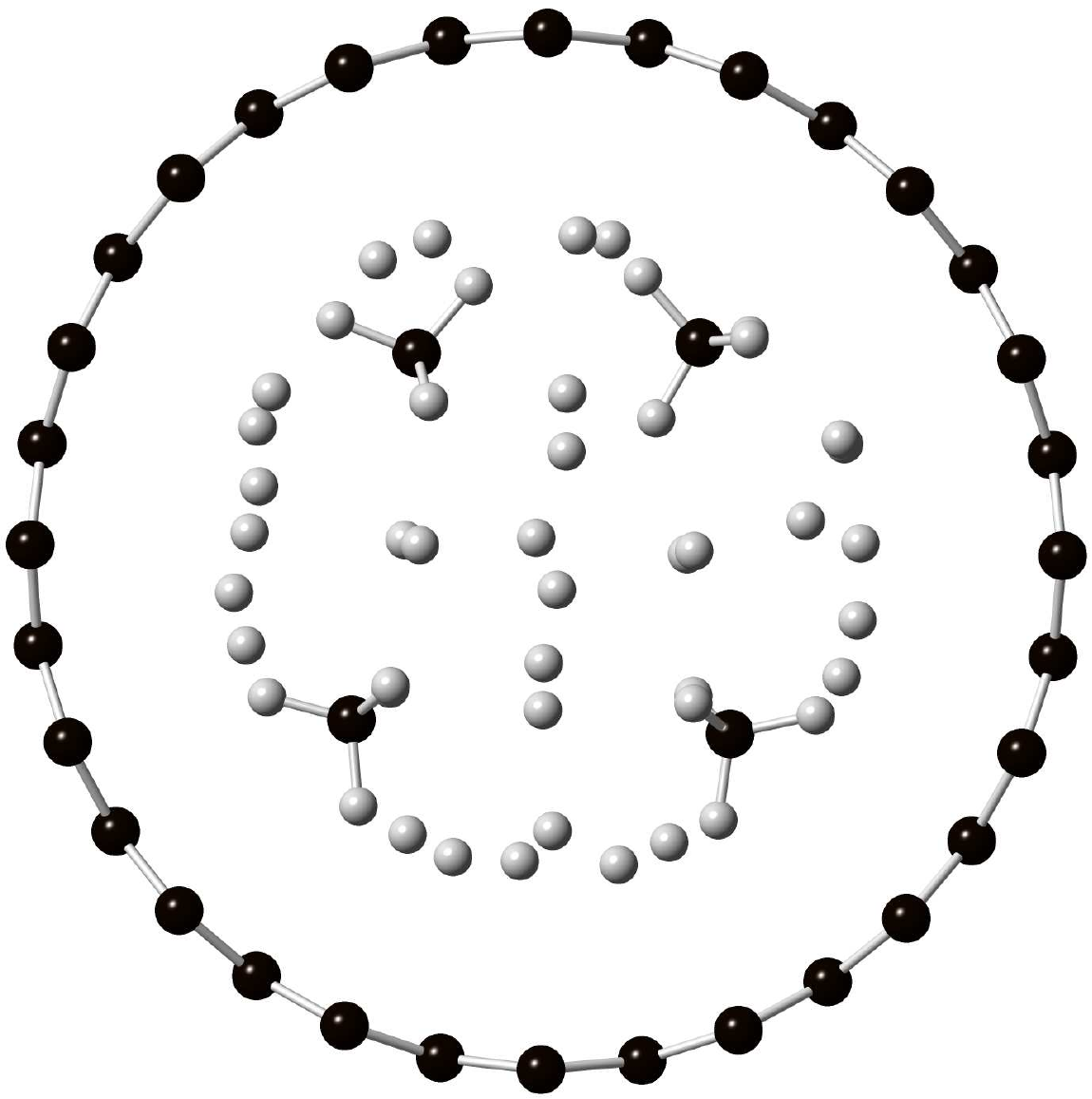}
\caption{\label{fig:H4M_tube}Optimized structure of a $2\times 2\times
1$ supercell of H4M inside a (16,0) carbon nanotube.}
\end{indented}
\end{figure}

The length of one of the edges of the SO H4M crystalline unit cell at
5.8~GPa is very close to the unit cell length of carbon nanotubes in
axial direction (4.27~\AA). Hence, we compress the H4M unit cell
slightly so that the edge is exactly commensurate and scale the other
two edges accordingly. We then put $n\times n\times 1$ supercells of H4M
inside the carbon nanotubes and relax the whole system. For example,
since the diagonal length of the scaled H4M unit cell is 6.6~\AA\
(slightly bigger than the radius of the (16,0) carbon nanotube), it is
natural to build the filled structure from a (16,0) nanotube together
with a $2\times 2\times 1$ H4M supercell. We slightly adjust the
position of several hydrogen molecules at the corners to prevent steric
clashes. The resulting relaxed structure is shown in
\fref{fig:H4M_tube}.  Throughout the relaxation, the structure of H4M
changes noticeably. The nanotube also deforms and exhibits a strain of
1.4\%, which is more than we estimated in \tref{tab:H4M-CNT} since we
compressed the supercell slightly to make it commensurate; but
nevertheless, it withstands the pressure created by the H4M inside.

In \tref{tab:H4M-CNT}, we list the theoretically possible hydrogen mass
density for the CNT+H4M system for nanotubes up to a radius of 25~\AA;
even better performance might be achievable for larger tubes. Note that
for tubes starting at (40,0), the filling with supercells of H4M is not
simply a square-type pattern corresponding to $n\times n\times 1$, but
we can fit in further unit cells on the side to better approximate the
circular shape of the tube. The corresponding filling is then reported
as $n\times n\times 1 + m$, where $m$ indicates the number of additional
unit cells of H4M that we are able to fit in a given tube. According to
our estimation, all indicated tubes provide enough pressure to stabilize
H4M at room temperature.

\begin{table}
\caption{\label{tab:H4M-CNT}Theoretically predicted performance of the
CNT+H4M systems. Given are the chirality of the system, the nanotube
radius [\AA], the largest possible H4M supercell commensurate with the
diameter, the mass densities $\rho_{\rm mass}$ (with and without
including the hydrogen in CH$_4$) [mass\%], the volumetric densities
$\rho_{\rm vol}$ [kg H$_2$/L], the radial Young's modulus $E_r$ [GPa],
and the strain $\varepsilon$ [\%] required to produce 5.8 GPa.}
\begin{indented}
\item[]\begin{tabular}{@{}lccccccr@{}}
\br
System & Radius & Filling & $\rho_{\rm mass}^{\rm w}$ & $\rho_{\rm mass}^{\rm w/o}$ & $\rho_{\rm vol}$ & $E_r$ & $\varepsilon$\\
\mr
(10,0) & 3.95  & $1\times 1\times 1$      & 2.4  & 1.6  & 0.028 & 870 & 0.7\\
(16,0) & 6.30  & $2\times 2\times 1$      & 5.6  & 3.7  & 0.055 & 555 & 1.0\\
(24,0) & 9.43  & $3\times 3\times 1$      & 7.9  & 5.3  & 0.065 & 373 & 1.6\\
(32,0) & 12.56 & $4\times 4\times 1$      & 10.0 & 6.7  & 0.070 & 280 & 2.1\\
(40,0) & 15.69 & $5\times 5\times 1$ + 8  & 14.6 & 9.7  & 0.097 & 224 & 2.6\\
(48,0) & 18.82 & $6\times 6\times 1$ + 12 & 16.7 & 11.1 & 0.102 & 187 & 3.1\\
(56,0) & 21.96 & $7\times 7\times 1$ + 16 & 18.4 & 12.2 & 0.103 & 160 & 3.6\\
(64,0) & 25.08 & $8\times 8\times 1$ + 24 & 20.4 & 13.6 & 0.110 & 140 & 4.1\\
\mr
H4M    & ---   & ---                      & 50.2 & 33.3 & 0.160 & --- & ---\\
\br
\end{tabular}
\end{indented}
\end{table}

\section{Conclusions}
\label{sec:conclusions}

In this paper we present results of \emph{ab initio} calculations on the
H4M system for the purpose of hydrogen storage. While H4M shows
exceptional hydrogen mass storage density---well beyond the required
Department of Energy target---it falls short in its thermodynamic
properties. It requires either very high pressure to be stable at room
temperature, or it needs to be extensively cooled at ambient pressure.
Our \emph{ab initio} simulations are a proof of concept that external
agents such as MOFs and carbon nanotubes can be used to provide the
necessary pressure. We find that certain MOFs provide enough pressure to
significantly decrease the burden of cooling, but none of them stabilize
H4M at room temperature. On the other hand, we find the very encouraging
result that carbon nanotubes may stabilize H4M at room temperature with
outstanding gravimetric and volumetric hydrogen storage densities.

While in this study we have performed basic simulations of possible host
materials to stabilize H4M, we have not addressed related practical
issues. Questions arise such as: even if these host materials are, in
principle, capable of stabilizing H4M at close-to-ambient temperatures,
how can we practically place the H4M inside the cavities of these host
materials? Also, it is not clear that the radial pressure of the
nanotubes alone is practically enough to stabilize H4M, some axial
pressure might be necessary. More research is needed to see if H4M can
be crystallized inside these structures, or if it can form clusters at a
low temperature and then diffuse into these cavities.

\ack

We would like to dedicate this report to the memory of \emph{Prof.\
David Langreth}, who passed away in mid-2011---he is the ``father'' of
vdW-DF and his research inspired many. All calculations were performed
on the WFU DEAC cluster.  This work was supported by the Department of
Energy Grant, Office of Basic Energy Sciences, Materials Sciences and
Engineering Division, Grant No. DE-FG02-08ER46491.


\end{document}